\documentclass[preprint,nofootinbib,longbibliography]{revtex4-1}
\usepackage[utf8]{inputenc}
\usepackage{amsmath}
\usepackage{amsfonts}
\usepackage{xcolor}
\usepackage{slashed}
\usepackage{mathtools}

\usepackage[colorlinks=true]{hyperref} 
\hypersetup{
    bookmarks=true,         
    unicode=false,          
    pdftoolbar=true,        
    pdfmenubar=true,        
    pdffitwindow=false,     
    pdfstartview={FitH},    
    pdfsubject={},   
    pdfcreator={},   
    pdfproducer={}, 
    pdfkeywords={} {} {}, 
    pdfnewwindow=true,      
    colorlinks=true,       
    linkcolor=magenta, 
    citecolor=blue,        
    filecolor=magenta,      
    urlcolor=blue           
}

\begin{document}
\title{Hall conductivity pump}
\begin{abstract}
The Thouless charge pump represents a transfer of electric charge through a gapped one-dimensional system between its zero-dimensional boundaries under a periodic change of a parameter. The value of the passed charged during a single cycle is known to be a topological invariant. We construct an analogous topological invariant that measures a pump of Hall conductance inside of three-dimensional material between its two-dimensional boundaries.
\end{abstract}
\author{Lev Spodyneiko}
\email{lionspo@mit.edu}
\affiliation{Massachusetts Institute of Technology, Cambridge, MA 02139, United States}
\maketitle

\section{Introduction}
The primary objective of the present paper is to build on previous work \cite{Berry,Thouless} and construct an invariant that measures the pumping of Hall conductance in three and higher dimensions. We will take the opportunity and review the physical idea of the construction.

This work is focused on topological invariants of gapped systems. The simplest example of that is a zero space dimensional (0d) system with finite Hilbert space which conserves $U(1)$ charge. If the ground state is separated by a gap from the rest of the spectrum the expectation value of the charge operator is an integer. The latter cannot change under continuous deformations of the Hamiltonian as long as the gap is preserved and the charge conservation is respected. In this sense, it is a topological invariant and this is what is meant by the term topological in the rest of the paper.

There is no straightforward generalization of this invariant to a one-dimensional (1d) system since the charge of the ground state typically scales intensively with the size of a system. Nevertheless, there is a more sophisticated generalization. One-dimensional systems always have zero-dimensional systems at the boundary which can have a non-trivial charge. However, one can always change the charge of the boundary by stacking it with a decoupled 0d system and thus it is inherently ambiguous. This is not very interesting and in order to get something non-trivial one can employ intuition from bulk-boundary correspondence. In the present setting, this would mean that the conservation of charge on the boundary is anomalous because the charge can flow into the bulk. An example of it is the Thouless charge pump \cite{TP,Niu} when the charge is transmitted from one boundary to the other through a gapped bulk by a periodic adiabatic change of a parameter.

The value of the pumped charge is a topological invariant of a family of systems in the following sense. Consider a family gapped 1d Hamiltonians parametrized by a periodic parameter. One can compute how much charge passes through a fixed section as the parameter is periodically varied.  The result will not change if we add a small perturbation to the family as long as each member of the family stays gapped.  The charge passing through a fixed section during a single cycle is a topological invariant that protects the family from deformation to a trivial family, i.e. a family that does not depend on the parameter. This way one can construct a topological invariant of a family from a topological invariant living in a lower dimension. 

Heuristically, one can imagine that a charged mode, localized on one boundary, delocalizes and moves through the bulk onto another boundary \cite{Poshen}.  A natural question is whether a similar thing can happen for other topological invariants. The answer is positive, and an example is a mode with a non-trivial Chern class of the Berry connection \cite{Xueda} (which is actually a topological invariant of a 2-parameter family of 0d systems) that can move through the bulk from one edge to another. This leads to an invariant of a 3-parameter family for 1d system.

Both of these examples are interesting but they start from invariants of 0d systems. It is interesting to find a generalization where a higher dimensional topological invariant is "pumped" through the bulk. One can pump a non-trivial family through the bulk. Namely, one can construct an invariant \cite{Thouless} that measures how many 1d Thouless pumps have been pumped from one 1d edge of 2d system to another. This is intriguing but somewhat unsatisfactory since the pumped quantity is an invariant of a family itself. 

In the present paper, we will construct a topological invariant that measures the pumping of a 2d Hall conductance inside of 3d material under periodic change of a parameter. More precisely, consider a family of gapped 3d systems parametrized by a periodic parameter with a boundary consisting of two planes.  The invariant measures by how much the quantum Hall conductance of one boundary changes during an adiabatic change of a parameter.

This characterization of the invariant contains an important caveat which is a manifestation of the bulk-boundary correspondence for topological invariants of families. The easiest way to explain it is to use an example of a Thouless charge pump. There are two ways to compute the pumped charge. One is to use the static perturbation theory and find a charge passing through a section of an infinite 1d system. The system is assumed to be gapped and to have a unique ground state. The parameter is assumed to be varied adiabatically and thus we can use the adiabatic theorem. On the other hand, one can consider a finite-size system with two boundaries and compute the change of charge on one of the boundaries. In this case, one can try to apply the adiabatic theorem but it leads to a problem.  The theorem states that the system should stay in the ground state. This would mean that the charge of each of the boundaries would return to the initial value after a complete cycle resulting in a zero passed charge. The seeming contradiction between the two computations lies in the applicability of the adiabatic theorem. The latter assumes that the instantaneous ground state does not intersect other states for all values of the parameters. This leads to the bulk boundary correspondence of invariants of the families: whenever the invariant computed in the bulk is non-trivial one cannot introduce a gapped boundary condition for all members of the family which continuously depend on the parameters.

A seemingly different way of how a topological invariant can be generalized to higher dimensions is via a charge of solitonic objects. For example, in 1d a soliton can carry a non-trivial charge which is protected under continuous deformations. The existence of charged defects can be used to characterize the topologically protected properties of a state.  However, this approach can be related to the previous one. Instead of considering slow adiabatic variation in time, one can consider slow variations in space. As one moves along the soliton, the parameters of the Hamiltonian change. Far away from a soliton the Hamiltonian should return to itself, thus one can think about the change of the parameter as one moves perpendicular to the soliton as a loop in the parameter space. The Thouless charge pump of this cycle is related to the charge of the soliton \cite{Thouless}. Similarly, one can think about domain wall defects in 3d system with non-trivial 2d Hall conductivity. 

A one-parameter family of systems represents a loop in the space of gapped theories $\mathfrak M_D$ in $D$-dimensions. Whenever a topological invariant is non-trivial the loop cannot be contracted to a point and represents a non-trivial element in the fundamental group $\pi_1 (\mathfrak M_D)$. Similarly, invariants of $n$-parameter families distinguish non-trivial elements of $\pi_n(\mathfrak M_D)$. Typically, one is only interested in different gapped phases that correspond to different connected components $\pi_0 (\mathfrak M_D)$ of  $\mathfrak M_D$. One might wonder why one should be interested in more sophisticated details of the topology of $\mathfrak M_D$ besides its zeroth homotopy group $\pi_0 (\mathfrak M_D)$.

First, as we discussed above, non-trivial families are related to solitons and other defects. Secondly, it is interesting that each member of the family can be in a trivial phase (i.e. it can be continuously deformed into a trivially gapped system) while the family as a whole is non-trivial. Lastly, in the case of the invertible topological orders, it is expected \cite{Kitaev_SRE,Freed} that $\mathfrak M_D$  forms a loop spectrum $\pi_n(\mathfrak M_D) =\pi_{n+1}(\mathfrak M_{D+1})$. The construction of invariants of families can be used to test and prove these conjectures.

The above discussions were very intuitive and gave a simple physical picture. However, it was heuristic at best and contained numerous implicit assumptions. Even in the case of the charge, separating the boundary from the bulk and tracking the mode is a very sophisticated procedure and is plagued with ambiguities.  For the Hall conductance pump, the matter is further complicated by pumping a subsystem of macroscopic size. We can avoid the technical issues by using the descent equation coined by Kitaev \cite{Kitaev_talk} and further developed in \cite{Berry,Thouless}. Instead of dealing with the pumps directly, it defines a form on the parameter space which can be computed by a localized static response. The integral of this form over the parameter space can be shown to be a topological invariant of the family and can be interpreted as a pump: a non-trivial flow of 2d system under a periodic adiabatic pump inside of 3d system. The descend equation can be applied further to construct invariants of $n$-parameter families in $D+n-2$ dimensions.  

Since Hall conductivity is not an integral of a localized density the formalism of the descent equation is not directly applicable to it. Fortunately, one can use St\v{r}eda formula in order to write Hall conductivity in the form appropriate for the descent equation. Even though the pumped quantity is two-dimensional in nature the resulting invariant is given by a formula that only has contributions localized around a point. 

In the present paper, we will use Hamiltonian formalism and by a family of systems, we will mean a family of gapped Hamiltonians. This allows expression for invariants in terms of standard perturbation theory. The resulting invariant only depends on the ground-state wave function with other details of the Hamiltonian being irrelevant. In the purely wave-functions language these results were derived in \cite{Noether}. Also, similar ideas in the setting of quantum field theory were studied in \cite{QFTpumps1,QFTpumps2}.

The paper is structured as follows. After explaining our setting in Section \ref{sec: lat sys}, we review the descend construction on the example of a charge pump in Section \ref{sec: thouless pump}. We derive the higher-dimensional invariants for the Hall effect in Section \ref{sec: Hall des}. Field theory interpretation as well as applications are discussed in Section \ref{sec: app}. We summarize the bookkeeping notations in Appendix \ref{app: notation} and derive the St\v{r}eda formula for lattice systems in Appendix \ref{app: streda}.

We thank   Anton Kapustin and Nikita Sopenko for the discussions. The work was supported by the Simons Collaboration on Ultra-Quantum Matter, which is a grant  (651446) from the Simons Foundation.

\section{Lattice systems}\label{sec: lat sys}
In this paper, we consider lattice systems defined on a subset $\Lambda \subset\mathbb R^d$. The total Hilbert space is a tensor product of onsite Hilbert spaces. The latter is assumed to be finite-dimensional. The closure of the lattice $\Lambda$ is assumed to be discrete. We will restrict to models which are invariant under $U(1)$ symmetry generated by charge densities $Q_p$ with $p\in \Lambda$. They are hermitian operators with integer eigenvalues such that
\begin{align}
    [Q_p,Q_q] &= 0,\\
    \sum_{q\in \Lambda}[Q_q,H_p] &=0,
\end{align}
where $H_p$ is Hamiltonian density. Operators $Q_p$ and $H_p$ are assumed to act trivially on a site $q$ if $|q-p|>R_{\rm int}$ for all $p,q$ and some constant $R_{\rm int}$. We will only consider gapped Hamiltonians with a unique ground state. 

The time-derivative of the charge density reads
\begin{align}
    \frac{d Q_p}{dt} = i[H,Q_p] = -\sum_{q\in\Lambda} J_{qp},
\end{align}
where we have defined the current as
\begin{align}
    J_{qp} = i[Q_p,H_q]-i[Q_q,H_p].
\end{align}
It represents the current flowing from $p$ to $q$ and is thus chosen to be anti-symmetric. Suppose, $f(p)$ is 1 on a finite subset $A\subset \Lambda$ and zero otherwise. Then, the charge in this region is
\begin{align}
    Q(f) = \sum_{p\in\Lambda} f(p)Q_p.
\end{align}
It evolves in time as
\begin{align}
    \frac{dQ(f) }{dt} = - J(\delta f),
\end{align}
where we have defined 
\begin{align}
    J(\delta f) = \frac 1 2 \sum_{p,q\in\Lambda} (f(q)-f(p)) J_{pq},
\end{align}
which physically corresponds to the current flowing through the boundary of $A$. We defined it for a function $f$ with compact support, but it can be generalized to include more general functions such as  $f(p)=\theta (a-x^i(p))$, where $x^i(p)$ is the  $i$th coordinate of the site $p$ in $\mathbb R^d$. In this case, $J(\delta f)$ will represent the total charge flowing through the hyperplane $x^i =a$.

In a ground state, the charge density is constant 
\begin{align} \label{eq: zero current}
   \sum_{p\in\Lambda}\langle  J_{pq} \rangle =0.
\end{align}
This equation can be solved
\begin{align}\label{eq: j  curl M}
    \langle  J_{pq} \rangle = \sum_{r\in\Lambda} M_{pqr},
\end{align}
where $M_{pqr}$ is magnetization density \cite{THE} which is totally antisymmetric in $p,q,r$. Solution of eq. (\ref{eq: zero current}) requires specifying boundary conditions and magnetization non-locally depends on them. However, the variation of magnetization with respect to change of parameters is local \cite{KitaevAnyons,THE,nernst}  and is given by

\begin{align}
    \frac {dM_{pqr}}{d\lambda} = \oint_{z=0} \frac{dz}{2\pi i}  \text{Tr}\left( G \frac{dH_p}{d\lambda} G J_{qr}+G \frac{dH_q}{d\lambda} G J_{rp}+G \frac{dH_r}{d\lambda} G J_{pq} \right),
\end{align}
where we introduced the resolvent $G= \dfrac 1 {z-H}$ to shorten the formulas. The integral is around ground state energy $z=E_0$ which we choose to be 0.

Later, we will need a derivative of magnetization with respect to the chemical potential $\mu$. It is related \cite{Luttinger}  to derivative of magnetization with respect to a deformation $dH_p=d \lambda_Q \,Q_p$ as $\dfrac{dM}{d\mu} = -\dfrac{dM}{d\lambda_Q} $ and is given by
\begin{align}\label{eq: M mu def}
    \frac {dM_{pqr}}{d\mu} =- \oint_{z=E_0} \frac{dz}{2\pi i}  \text{Tr}\left( G Q_p G J_{qr}+G Q_q G J_{rp}+G Q_r G J_{pq} \right).
\end{align}

\section{Descendants from Thouless pump.} \label{sec: thouless pump}

In this section, we explain the basic idea behind the construction of higher dimensional topological invariants. We focus on the example of a first descendant of 0d electric charge of the ground state. The result is a Thouless charge pump. This example is simple and physically transparent while highlighting the main ideas of the construction. 

Suppose for a moment that the number of cites in $\Lambda$ is finite. Then one has a finite-dimensional quantum mechanical system. Since the Hamiltonian is gapped and has a unique ground state the expectation value of the total charge in the ground state $\langle Q\rangle $ is an integer. It remains unchanged under continuous deformations of Hamiltonian as long as it stays gapped and commutes with the charge operator. So, it is a topological invariant of 0d systems. 

If one considers a one-dimensional lattice the charge of the ground state will typically diverge as volume increases and it is ill-defined in infinite volume. However, one can expect the charge passing through any section under a deformation of the Hamiltonian to be finite.  

Consider a function $f(x)= \theta(x-a)-\theta(x-a-L)$, i.e. $f(x)$ is 1 if $a<x<a+L$ and zero otherwise. We assume that $a,a+L$ are not points of 1d lattice $\Lambda\subset \mathbb R$. Consider an infinitesimal deformation of a system given by $H_p\rightarrow H_p+dH_p$. The change of the ground state expectation value of the total charge in the region $a<x<a+L$ is 
\begin{align}
 \sum_{a<p<a+L}d   \langle Q_p\rangle  = \sum_{p\in \Lambda} f(p)d \langle Q_p\rangle =-\sum_{p,q\in \Lambda} f(p)\left( \langle dH_q P\frac 1 H P Q_p \rangle + \langle Q_p P\frac 1 H P  dH_q\rangle \right),
\end{align}
where $P$ is the projector onto the excited states and we used the quantum mechanical perturbation theory. We normalized the ground state energy to be zero. Each summand in the last formula is non-zero when $p,q$ are deep inside the region $a<x<a+L$. However, physically we would expect the flow of the charge to occur only at the boundaries around $a$ and $a+L$. This calls for the following rearrangement of the summation
\begin{align}
\begin{split}
    \sum_{p,q\in \Lambda} f(p)\langle dH_q P\frac 1 H P Q_p \rangle  &= \sum_{p,q\in \Lambda} f(p)\left(\langle dH_q P\frac 1 H P Q_p \rangle -\langle dH_p P\frac 1 H P Q_q \rangle\right) \\&= \frac 1 2 \sum_{p,q\in \Lambda} (f(p)-f(q))\left(\langle dH_q P\frac 1 H P Q_p \rangle -\langle dH_p P\frac 1 H P Q_q \rangle\right).
\end{split}
\end{align}
In the first equality, we used the fact that the total charge commutes with the Hamiltonian $H$ and therefore with the projection $P$.  The projection acting on the ground state gives zero. In the second equality, we used the antisymmetry in $p,q$ of the expression in the brackets. Doing the same to the other term, we find
\begin{align}
    \sum_{p\in \Lambda} f(p)d \langle Q_p\rangle = \frac 1 2 \sum_{p,q\in \Lambda} (f(q)-f(p))Q^{(1)}_{pq} = Q^{(1)}(\delta f),
\end{align}
where we have defined 
\begin{align}
   Q^{(1)}_{pq}&= \langle dH_q P\frac 1 H P Q_p \rangle-\langle dH_p P\frac 1 H P Q_q\rangle+\langle Q_p P\frac 1 H P dH_q \rangle-\langle Q_q P\frac 1 H P dH_p\rangle,\\
   Q^{(1)}(\delta f)&=\frac 1 2 \sum_{p,q\in \Lambda} (f(q)-f(p))Q^{(1)}_{pq}.
\end{align}
The function $Q^{(1)}_{pq}$ rapidly decays \cite{Watanabe} to zero as $|p-q|\rightarrow \infty$. After these manipulations, one can write
\begin{align}
 d \sum_{a<p<a+L}\langle Q_p\rangle =  Q^{(1)}(\delta f) \approx Q^{(1)}(\delta f^L)-Q^{(1)}(\delta f^R),
\end{align}
where $f^L (x)=\theta(x-a)$ and $f^R (x)=\theta(x-a-L)$. The error is of order $O(L^{-\infty})$.

A couple of important points. First, even though the charge 
\begin{equation}
    Q^{(0)}(g)=\sum_{p} g(p) \langle Q_p\rangle
\end{equation} 
is ill-defined for $g=f^{L,R}$, the amount of the charge $Q^{(1)}(\delta f_{L,R})$ flowing through each edge of the region under deformation $dH$ is well-defined. Second, the total change of the charge $Q^{(0)}(f)$ as a parameter of Hamiltonian is periodically varied is zero. However, suppose we only focus on one boundary and compute the integral $\oint_{S_1} Q^{(1)}(f^L)$ along the cycle in the parameter space. Here the integral is over a periodic parameter of the Hamiltonian $\lambda\sim \lambda+1$. In that case, the result may be non-trivial and is actually a topological invariant in the sense that we will explain momentarily. This is a Thouless charge pump. It measures how much charge has flown under the adiabatic periodic change of a parameter. The charge passing during one cycle cannot change under continuous variations of the Hamiltonian as long as it stays gapped.

The main idea behind descend construction is to extend this procedure by replacing the charge with other topological invariants. For example, one can start from the Thouless pump $\oint Q^{(1)}(f^L)$ instead of the charge $\langle Q\rangle$. Above we used the charge density $Q^{(0)}_p=\langle Q_p\rangle$ extensively, an analog for the pump would be $ \frac 1 2\sum_{q \in \Lambda}\oint Q^{(1)}_{pq} (f^L(q)-f^L(p))$. The last expression looks cumbersome due to parameter space integration as well as a complicated contraction with the function $f$. In order to simplify the computation, it is easier to work with densities valued in differential forms on the parameter space and do the integration in the parameter space and contraction with functions at the very last step. 

The essential step in the construction of the Thouless pump is to find the solution $Q^{(1)}_{qp}$ of the equation
\begin{align}
    dQ_p^{(0)}=d\langle  Q_p \rangle = \sum_{q\in\Lambda} Q^{(1)}_{qp}.
\end{align}
The function $Q^{(1)}_{qp}$ must be chosen antisymmetric in $p,q$ in order to have zero contribution from the region where $f$ is constant. This equation can be generalized to\footnote{One should note that there are spurious non-trivial solutions to the equation $\sum_{q\in\Lambda}A_{q,p_0,\dots,p_n}^{(n+1)}=0$. An example in 1d with the lattice $\Lambda = \mathbb Z$ is $A_{p,q}= \delta_{p,q+1} - \delta_{p+1,q}$. In order to exclude such solutions we impose additional requirements. First, $Q^{(n)}$ must be linear in deformation, i.e. $Q^{(n)}$ for $dH= dH_1 +dH_2$ is a sum for each of them. Secondly, if we replace the  deformation $dH_p$ with $\theta(|p|-R)dH_p$ then $Q^{(n)}_{p_0,\dots p_n}$ must go to zero as $R\rightarrow \infty$ with $p_i$ being fixed. In other words, $Q^{(n)}_{p_0,\dots p_n}$ should go to zero not only when two of $p_i$ are well separated, but also when one of the points is far away from the support of the deformation.}

\begin{align}\label{eq: descendant simple form}
    dQ_{p_0,\dots,p_n}^{(n)} = \sum_{q\in\Lambda}Q_{q,p_0,\dots,p_n}^{(n+1)},
\end{align}
where $Q_{q,p_0,\dots,p_n}^{(n+1)}$ is $(n+1)$-form on the parameter space which is antisymmetric in the indices $q,p_0,\dots,p_n$. The 
exterior derivative $d=\sum_i d\lambda_i\dfrac \partial{\partial \lambda_i}$ acts on the parameters $\lambda_i$ of the Hamiltonian.  

As one goes to higher dimensions it is getting harder to keep track of the different indices. Fortunately, a convenient formal language reviewed in Appendix \ref{app: notation} drastically simplifies the formulas as well as makes the interpretation clearer. In the following, we will use these notations.  The descendant equation (\ref{eq: descendant simple form}) takes the form
\begin{align}
d Q^{(n)} = \partial Q^{(n+1)}.
\end{align}
The solution to this equation can be found in \cite{Thouless}.

By evaluating $n$-chain  $Q^{(n)}$  on $n$-cochain $\alpha^{(n)}$, we get an $n$-form $\langle Q^{(n)},\alpha^{(n)}\rangle$ on the parameter space.

If cochain $\alpha^{(n)}$ is closed $\delta \alpha^{(n)} =0$, then the resulting form will be closed with respect to~$d$
\begin{align}
    d \langle  Q^{(n)}, \alpha^{(n)}\rangle = \langle  \partial Q^{(n+1)}, \alpha^{(n)}\rangle = \langle  Q^{(n+1)}, \delta\alpha^{(n)}\rangle =0.
\end{align}
The result integration of $n$-form $\langle  Q^{(n)}, \alpha^{(n)}\rangle$ over a manifold without boundary in the parameter space will not depend on continuous deformations of this manifold:
\begin{equation}
   \int_M \langle  Q^{(n)}, \alpha^{(n)}\rangle -  \int_{M'} \langle  Q^{(n)}, \alpha^{(n)}\rangle =  \int_{\partial Y} \langle  Q^{(n)}, \alpha^{(n)}\rangle =   \int_Y d\langle  Q^{(n)}, \alpha^{(n)}\rangle =0,
\end{equation}
where $M'$ is the result of a continuous deformation of $M$ and $Y$ is $(n+1)$-dimensional manifold which $M$ sweeps as it gets continously deformed into $M'$. The boundary of $Y$ consists of $M$ and $M'$. For closed $\alpha$ and $M$ without a boundary, we will call $\int_M \langle  Q^{(n)}, \alpha^{(n)}\rangle$ a topological invariant of a family of gapped Hamiltonians parameterized by $M$. As one can see, it does not change under continuous deformations of $M$.  If $M$ can be continuously contracted into a point, the resulting invariant will be zero. 

On the other hand, if $\alpha^{(n)}$ is exact $\alpha^{(n)}=\delta \gamma^{(n-1)}$, we find
\begin{align}
     \langle  Q^{(n)}, \alpha^{(n)}\rangle = \langle   Q^{(n)}, \delta \gamma^{(n-1)}\rangle = \langle \partial Q^{(n)}, \gamma^{(n-1)}\rangle =d\langle Q^{(n-1)}, \gamma^{(n-1)}\rangle.
\end{align}
In this case, the integration over a manifold without boundary  gives zero
\begin{align}
    \int_M   \langle  Q^{(n)}, \alpha^{(n)}\rangle = \int_M   d\langle  Q^{(n-1)}, \gamma^{(n-1)}\rangle = \int_{\partial M}   \langle  Q^{(n-1)}, \gamma^{(n-1)}\rangle=0.
\end{align}
 Thus, the topological invariant  $\int_M \langle  Q^{(n)}, \alpha^{(n)}\rangle$ is trivial if $\alpha^{(n)}$ is exact. 

From the above, we see that if one evaluates $Q^{(n)}$ on a non-trivial element of $n$-cochain cohomology and integrates the result over a manifold without boundary in the parameter space one would get a topological invariant of the family of gapped Hamiltonians. 

This naturally leads us to the question of what is chain cohomology of the lattice $\Lambda \subset \mathbb R^{D}$. For lattices that are coarsely equivalent to $\mathbb Z^{D}$ (roughly, lattices which can be deformed into $\mathbb Z^D$ by a finite distance shift of every site such that only a finite number of sites accumulate at every point), one only has a non-trivial element in $D$-cochain cohomology. This means that one will construct a $D$-form $\langle  Q^{(D)}, \alpha^{(D)}\rangle $  for $D$-dimensional family of Hamiltonians.  

In the same way, one can construct \cite{Berry}  higher Berry curvatures $\Omega^{(n)}$, which physically correspond to Thoughless pumps of the Chern class of the Berry curvature.  In the next section,  we will construct  descendants $H^{(n)}$ of the Hall conductance. 
\section{Descendants of Hall effect}\label{sec: Hall des}

The electric Hall effect in two-dimensional space at zero temperature is given by \cite{THE}
\begin{equation}
\sigma^H(f,g)=  -i\oint_{z=E_0} \frac{dz}{2\pi i} \text{Tr}\left( G J(\delta f) G^2 J(\delta g) \right),
\end{equation}
where $f= \theta(-x^1)$ and $g= \theta(-x^2)$. It is a topological invariant and we would like to construct descendants of it. However, after striping it off from the contraction with functions $f,g$ we find
\begin{equation}
    -i\oint_{z=E_0} \frac{dz}{2\pi i} \text{Tr}\left( G J_{pq} G^2 J_{rs} \right),
\end{equation}
which is not totally antisymmetric in $p,q,r,s$. Moreover, it has 4 indices and in 2d we expect it to have only 3 since only 2-cochains cohomology is non-trivial in this dimension. The way to cure both of these problems is to use the St\v{r}eda formula
\begin{equation}\label{eq: streda formula}
 \sigma^H(f,g)=H^{(0)}(\delta f \cup \delta g),
\end{equation}
where 2-chain $H^{(0)}$ is minus derivative of magnetization with respect to the chemical potential given by 
\begin{align}\label{eq: H0 def}
H^{(0)}_{pqr} =  - \frac {dM_{pqr}}{d\mu} = \oint_{z=E_0} \frac{dz}{2\pi i}  \text{Tr}\left( G Q_p G J_{qr}+G Q_q G J_{rp}+G Q_r G J_{pq} \right).
\end{align}

The equation (\ref{eq: streda formula}) is the lattice analog of the St\v{r}eda formula \cite{Streda1,Streda2} in the continuum which reads
\begin{align}\label{eq: streda continuum}
    \sigma^H =  \left( \frac{\partial n}{\partial B}\right)_{\mu} = \left( \frac{\partial M_z}{\partial \mu}\right)_{B},
\end{align}
where $n,B,M_z,\mu$ are density, magnetic field, magnetization, and chemical potential respectively. We prove (\ref{eq: streda formula}) in the appendix \ref{app: streda}. The extra minus sign in (\ref{eq: H0 def}) compared to (\ref{eq: streda continuum}) comes from the fact that the continuum limit of equation (\ref{eq: j  curl M}) is 
\begin{align}
    J_i({\bf x}) = -\epsilon_{ij}\partial_j M({\bf x}),
\end{align}
and the lattice definition of magnetization (\ref{eq: j  curl M}) is related to the continuum one as $M=-M_z$.

Now, the Hall conductance is in an appropriate form for the construction of descendants. The solution to the descendant equation
\begin{align}
    dH^{(n)} = \partial H^{(n+1)}
\end{align}
is
\begin{align}
\begin{split}
    H^{(n)}_{p_0,\dots,p_n} = \frac 1 2 \sum_{\sigma \in S_{n+2}} (-1)^{{\rm sgn} \,\sigma} \oint \frac {dz} {2\pi i } \sum_{j=0}^n (-1)^{n-j}A^{(j)}_{p_{\sigma(0)},\dots,p_{\sigma(n+2)}},
\end{split}
\end{align}
where 
\begin{equation}
    A^{(j)}_{p_0,\dots,p_{n+2}}  = {\rm Tr} \Bigg ( G dH_{p_0}GdH_{p_1} \dots GdH_{p_{j-1}}G Q_{p_j}GdH_{p_{j+1}} \dots  G dH_{p_n} G J_{p_{n+1},p_{n+2}}\Bigg).
\end{equation}
The first sum is over permutations of $n+2$ elements. We omitted the wedge product of forms (on the parameter space) from the last formula. Using results of \cite{Watanabe}, one can show that $H^{(0)}_{pqr}$ decays to zero fast when either of $|p-q|,|q-r|,|r-p|$ is large. We expect that similar results hold for other $H^{(n)}$ and they are chains in the sense of Appendix \ref{app: notation}. In the following, we will assume this.   

The $n$-chain $H^{(n)}$ has to be contracted with the only non-trivial element of the cohomology   and integrated over $(D-2)$-dimensional manifold  without boundary  $M$ in the parameter space leading to a topological invariant of a family

\begin{align}\label{eq: higher hall contructed}
  \int_M  H^{(D-2)}(\alpha^{(D)}).
\end{align}

A simple non-trivial example of such a co-chain is 
\begin{align}
    \alpha^{(D)} = \delta f_1 \cup \delta f_2\cup \dots\cup  \delta f_{D},
\end{align}
where $f_i = \theta(x^i(p) - a_i)$ with $x^i(p)$ being the $i$th coordinate of site $p$. Due to the fast decay of correlation functions in the definition of $H^{(D-2)}$ the resulting invariant will be localized around the point given by coordinates $(a_1,\dots, a_D)$. Note, that different choices of the point   $(a_1,\dots, a_D)$ can be related to each other by the addition of exact one-form to $\delta f_1 \cup \dots\cup  \delta f_{D}$, and thus the invariant does not depend on the choice of it. 

Suppose the topological invariant of a family is non-zero. It forbids boundary conditions that continuously depend on parameters for the whole family and preserve the gap.  Indeed, if the gap was preserved then the invariant would be the same on both sides of the boundary. However, it is zero on one side and non-zero on the other.

For a general gapped system the invariants (\ref{eq: higher hall contructed}) are not expected to be integer quantized. For invertible phases, i.e. the ones that can be deformed to a trivially decoupled phase after stacking it with an appropriate system, one can show that (\ref{eq: higher hall contructed})
are quantized as $\frac 1 {2\pi}$ times integer along the lines of \cite{Berry,Thouless} or more rigorous \cite{Artymowicz}. 

\section{Discussion and applications} \label{sec: app}

In this section, we will discuss some applications and properties of the invariants. Most of the discussion can be done within the framework of lattice systems but we will focus on field theory interpretation of the invariants.

 Whenever a gapped system admits a long-distance effective description in terms of field theory
  it lacks local low-energy degrees of freedom while the high-energy ones can be integrated out. The effective field theory still can capture interesting effects via its dependence on the background and by considering more sophisticated backgrounds one gains more insights into the underlying system. It is common to consider curved manifold $X$ with non-trivial metrics instead of flat space-time as well as turn on $U(1)$ background gauge field $A$ whenever one has local on-sight $U(1)$ symmetry. When the underlying system depends on parameters, one can make them space-time depend. If the parameter variations are smooth they would correspond to background fields $\phi: X\rightarrow M$ in the effective description. Here $M$ is the space of possible values for parameters and $\phi$ is a map that specifies their local values. Topologically protected quantities manifest themself in the form of topological terms in the effective action. The characteristic feature of the latter is that they are independent of the background metrics. In the following, we will discuss topological terms corresponding to the invariants of families in trivial phases. When non-trivial topological order is present one should include appropriate topological quantum field theories as well as their coupling to the background.
 
We first review the effective action for the Thouless charge pump~\cite{Thouless}. The 1d Thouless charge pump can be described by the following term in effective action
\begin{equation}\label{eq: QFT TP}
    S_{\rm top} (X,\phi,A) = \int_X  A\wedge \phi^*\tau^{(1)} = \int_X \epsilon^{\mu\nu} A_\mu \partial_\nu \phi^i \tau_i^{(1)}(\phi) d^2 x.  
\end{equation}
Here $\epsilon$ is a totally antisymmetric tensor, $x^0,x^1$ are time and space coordinates, and $\tau^{(1)}$ is 1-form on the parameter space $M$. In the classic example of adiabatic pumping via flux insertion into a cylinder, $M=S^1$ parameterizes the flux mod $2\pi$ and $\tau^{(1)}$ is properly normalized volume form on the circle. The relation between the Thouless pump and action (\ref{eq: QFT TP}) can be seen by computing the correction it introduces to the $U(1)$ current
\begin{align}\label{eq: QFT current TP}
    j^{\mu}_{\rm top}(x^0,x^1) = \epsilon^{\mu\nu}  \tau_i^{(1)}(\phi(x^0,x^1)) \partial_\nu \phi^i(x^0,x^1).
\end{align}
If we let the parameter depend only on time, and compute the charge flowing through a section $x^1=a$ during a single cycle, we find
\begin{align*}
    \Delta Q (a) = \int_0^T  j^{1}_{\rm top}(x^0,a) dx^0 = -\int_0^T   \tau_i^{(1)}(\phi(x^0)) \partial_0 \phi^i(x^0) dx^0 = -\int_{\phi(0)}^{\phi(T)}   \tau_i^{(1)}(\phi)  d\phi^i = -\oint_{S_1} \tau^{(1)},
\end{align*}
where we assumed that the parameters return to the initial values after a full cycle. The latter integral is the Thouless charge pump invariant. 

Instead of varying the parameter in the time one can vary it in space. Namely, consider $\phi^i(x^1)$ such that $\phi^i(-\infty)$ is the same point in $M$ as $\phi^i(\infty)$. If $\phi^i(x^1)$ sweeps a non-contractable circle in the parameter space $M$, this configuration can be interpreted as a soliton. The charge of this soliton can be found from (\ref{eq: QFT current TP})
\begin{align}
    Q_{\rm s} = \int  j^{0}_{\rm top}(x^1) dx^1 = \int \tau_i^{(1)}(\phi(x^1)) \partial_1 \phi^i(x^1) dx^1 = \oint_{S_1} \tau_i^{(1)}.
\end{align}
In higher dimensions, the Thouless pump topological term is
\begin{equation}
    S_{\rm top} (X,\phi,A) = \int_X  A\wedge \phi^*\tau^{(D)}, 
\end{equation}
where now $X$ is $(D+1)$-dimensional space-time and $\tau^{(D)}$ is a $D$-form on the parameter space. In the same way as above, one can find the correction to the current and show that this term represents the higher descendant of the Thouless charge pump.  Also, it gives charge to skyrmion in the same way it did to soliton. The relation between soliton/skyrmion charges and descendants of the Thouless pump can be seen directly on the lattice \cite{Thouless} without appeal to the effective description.   

We conclude the review of the Thouless pump by indicating the following relation with the $\theta$-terms. In 0d, there is only one topological action which is linear in vector potential
\begin{align}
   S= q\int_X A.
\end{align}
Gauge invariance forces the charge $q$ to be quantized and it is a topological invariant as we discussed at the beginning of Section \ref{sec: thouless pump}.  Are there any topological terms in 1d that a linear in $A$? There is a theta-term
\begin{align}\label{eq: theta term}
     S= \frac \theta {2\pi}\int_X F.
\end{align}
However, the coefficient $\theta$ is no longer forced to be quantized. It is a continuous parameter and one can deform it to zero without closing the gap. So it does not lead to a topological invariant in the usual sense. However, we can consider a space-time dependent theta-term
\begin{align}\label{eq: variable theta action}
     S= \frac 1 {2\pi}\int  \theta F = \frac 1 {2\pi}\int  \theta dA  = \frac 1 {2\pi}\int A \wedge d \theta,
\end{align}
which has the same form as (\ref{eq: QFT TP}). The theta term is periodic $\theta \sim \theta + 2\pi$ and thus naturally corresponds to a non-trivial loop in the parameter space. Contrary to the theta-term action (\ref{eq: theta term}) the coefficient in front of (\ref{eq: variable theta action}) is quantized and represents a topological invariant of a family.

Let us now turn to the descendants of Hall conductivity. The relevant term in the effective action is 
\begin{align}
\begin{split}
     S_{\rm top} (X,\phi,A) &= \frac 1 2 \int_X  A\wedge dA\wedge \phi^*h^{(D-2)} \\&= \frac 1 2 \int_X\epsilon^{\mu_1,\dots,\mu_{D+1}} A_{\mu_1} \partial_{\mu_2} A_{\mu_3} \partial_{\mu_3} \phi^{i_3} \dots \partial_{\mu_{D+1}} \phi^{{i_{D+1}}}h_{i_{3},\dots,i_{D+1}} ,
\end{split}
\end{align}
where $h^{(D-2)}$ is the $(D-2)$-form on the parameter space. For clarity, we will focus on the simplest case $D=3$.
\begin{align}
     S_{\rm top} (X,\phi,A) = \frac 1 2 \int_X  A\wedge dA\wedge \phi^*h^{(1)} = \frac 1 2 \int_X\epsilon^{\mu\nu\rho\sigma} A_{\mu} \partial_{\nu} A_{\rho} \partial_{\sigma} \phi^{i}h_{i}^{(1)}.
\end{align}
The correction to the current reads
\begin{equation}\label{eq: current HP}
    j^\mu_{\rm top} =  \epsilon^{\mu\nu\rho\sigma}\partial_{\nu} A_{\rho} \partial_{\sigma} \phi^{i}h_{i}^{(1)}.
\end{equation}
Heuristically, it can be related to the adiabatic pump of Hall conductivity in the following way.
 Consider a parameter that depends only on time and $A$ corresponding to a constant magnetic field $B_1$ along $x^1$ axis. The 2d charge denisty flowing through a section $x^1 =a $ is
 \begin{align}\label{eq: QFT QHE pump}
     \Delta n(a,B_1) = \int_0^T dx^0 \,j^1 = \int_0^T dx^0  \,B_{1}   \partial_{0} \phi^{i}h_{i}^{(1)} = B_{1} \oint h^{(1)}.
 \end{align}
Now, imagine that we have two boundaries at $x^1=a-L$ and $x^1=a+L$ then we have two 2d systems with $x^1<a$ and $x^1>a$ respectively.  Formula (\ref{eq: QFT QHE pump}) means that particle density of, say, the right subsystem will change by $\Delta n(a,B_1)$. Using St\v{r}eda formula (\ref{eq: streda continuum}) we find that the Hall conductivities of it change by $\oint h$ as expected.

Consider spatially dependent parameter $\phi(x^3)$, which depends only on the third coordinate and represents a non-trivial loop in the parameter space as $x^3$ changes from $-\infty$ to $\infty$. This represents a domain wall. Choose $A$ to represent constant electric field $E_2$ along $x^2$. Correction to the current (\ref{eq: current HP}) will give 
\begin{align}
   \int dx^3 j^2 = E_3\oint h^{(1)}.
\end{align}
This corresponds to the increase in 2d Hall conductance of the whole system by $\oint h^{(1)}$ due to the presence of the domain wall.

Descendants of the Berry curvature and the $U(1)$ charge give $D+2$ and $D$ forms on the parameter space respectively. The degree of these forms is greater or equal then the dimension of space. The descendant of the Hall conductance on the other hand gives a $(D-2)$-form and it can be used to detect gapless modes on defects. Let us illustrate this in the simplest case $D=3$. Suppose we have a line defect, such that the Hamiltonian has the density $H_p(\lambda(\phi))$ far away from the defect where $\phi$ is the angular coordinate around the defect. Far away from the defect, the system looks like a soliton from the previous paragraph. If the relevant integral $\oint_{S_1} H^{(1)}$ over $\lambda(\phi)$ for $\phi$ from 0 to $2\pi$ is non-zero this soliton will have a nontrivial Hall conductance.  The defect itself can be thought of as a boundary of this soliton. The non-zero Hall conductance of the soliton is known to be related to the level of Kac-Moody algebra at its boundary. Thus, we see that  $2\pi\oint_{S_1} H^{(1)}$ gives the level of the Kac-Moody algebra of the defect. 

Let us conclude with a couple of applications and specific examples. One can construct a translationally invariant free fermion system that has a non-trivial Hall conductivity pump $H^{(1)}$. The 1-form in this case can be related to the integral of the second Chern class of Bloch-Berry connection over 
3d Brillouin zone. The computations of the 1-form as well as an example of a system with a non-trivial second Chern class are analogous to the ones in \cite{Berry,Thouless} (see also \cite{Leung_2020}).

As an application of the invariant, one can consider a translationally invariant 3d system. There are 3 natural loops in the parameter space corresponding to shift by periods of the lattice in three directions. One can integrate one-form $H^{(1)}$ over these 3 loops and it will result in 3 components of invariants $\frac 1 {2\pi}\bf G$ of 3d quantum Hall state, where $\bf G$ is a vector of the reciprocal lattice. It cannot change unless either gap closes or translational symmetry is lost. Interestingly, dislocation characterized by Burgers vector $\bf B$ will host massless modes with level ${\bf B}\cdot {\bf G}$ Kac-Moody algebra \cite{TeoKane}. The descend equation in the presence of translational symmetry requires some care and we leave it to future work. There is an interesting question what happens in the case when there is a rotational symmetry on top of translational symmetry. In this case, dislocation can be split into two disclinations and it would be interesting to understand what happens to the massless modes.  
 
 \appendix
\section{Notation} \label{app: notation}
A large number of indices forces us to introduce an auxiliary formal definition in order to simplify the notation. See \cite{THE} for a more thorough discussion. 

We will call an antisymmetric function $A_{p_0,\dots,p_n}$ of $n+1$ lattice sites, which decays faster than any power of distance away from the diagonal $p_0=p_1=\dots,=p_n$, an $n$-chain.  From an $n$-chain one can construct an $(n-1)$-chain
\begin{align}
    (\partial A)_{p_1,\dots,p_n} = \sum_{p_0 \in \Lambda} A_{p_0,p_1,\dots,p_n},
\end{align}
where fast decay guarantees that the sum is well-defined. The boundary operator $\partial$ satisfies $\partial^2 =0$. For an antisymmetric function $f(p_0,\dots,p_n)$ (which do not necessarily decay away from the diagonal), we can define the contraction
\begin{align}
    \langle A,f\rangle = \sum_{p_0,\dots,p_n\in \Lambda} A_{p_0,\dots,p_n}f(p_0,\dots,p_n).
\end{align}
 In order for the sum to be well-defined one has to impose some conditions on $f$. In this paper, we will for simplicity consider only functions $f$ which are constant whenever all $|p_i-p_j|$ is greater than some fixed distance for every $i\ne j$. We will call such function $n$-cochain. Occasionally, we will write $A(f)$ instead of   $\langle A,f\rangle$.

One can define the co-boundary operator $\delta$ as
\begin{align}
    \delta f (p_0,\dots,p_{n+1}) = \sum_{j=0}^n(-1)^j f(p_0,\dots \hat p_j,\dots p_{n+1}),
\end{align}
where hat indicates the omitted variable. It is related to $\partial$ as 
\begin{align}
      \langle A, \delta f\rangle  = \langle \partial A,f\rangle.
\end{align}

A cup product of $n$-cochain with $m$-cochain which gives a $(n+m)$-cochain  defined as
\begin{align}
    f\cup g (p_0,\dots,p_{n+m}) = \frac 1 {(n+m+1)!} \sum_{\sigma \in S_{n+m+1}} (-1)^{{\rm sgn}\, \sigma} f(p_{\sigma(0)},\dots p_{\sigma(n)})g(p_{\sigma(n)},\dots p_{\sigma(n+m+1)}),
\end{align}
where the sum is over the permutation group. It satisfies
\begin{align}
    f \cup g &=(-1)^{|f||g|} g \cup f ,\\
    \delta(f\cup g) &= f \cup \delta g + (-1)^{|f|} f \cup \delta g,
\end{align}
where $|f|$ is the degree of the cochain.

We  define a cap product that gives an $n$-chain from $(n+m)$-chain and $m$-cochain
\begin{align}
\begin{split}
    A\cap f(p_0,\dots,p_n)&=(-1)^{mn}\frac{n!}{(n+m+1)!}\\&\times\sum_{p_{n+1},\dots,p_{n+m}}A(p_0,\dots,p_{n+m}) \sum_{i=0}^n f(p_i,p_{n+1},p_{n+2},\dots,p_{n+m}).
\end{split}
\end{align}
It satisfies
\begin{align}
    (A\cap f) (g) &= A(f \cup g),\\
    \partial(A\cap f) &= A (f), \qquad \text{when $|f|=|A|$}, \label{eq: normalization}\\
    \partial(A\cap f)&=(-1)^{|f|}\Big(\partial A\cap f - A\cap\delta f\Big).
\end{align}

Note that the cup product is defined even if $f$ is not a cochain, i.e. if it is not a constant when all its arguments are separated.

\section{Proof of St\v{r}eda formula} \label{app: streda}
In this appendix, we  show that the Hall conductivity
\begin{equation}
\sigma^H(f,g)=  -i\oint_{z=E_0} \frac{dz}{2\pi i} \text{Tr}\left( G J(\delta f) G^2 J(\delta g) \right).
\end{equation}
coincides with a minus derivative of the charge magnetization with respect to the chemical potential
\begin{align}
    H^{(0)} (\delta f \cup \delta g),
\end{align}
where 2-chain  $ H^{(0)}$ is given by
\begin{align}
    H^{(0)} = \oint_{z=E_0} \frac{dz}{2\pi i}  \text{Tr}\left( G Q_p G J_{qr}+G Q_q G J_{rp}+G Q_r G J_{pq} \right).
\end{align}
The idea behind the proof is to show that 1-chain
\begin{align}
    \sigma^H(\eta)_{pq} = -i\oint_{z=E_0} \frac{dz}{2\pi i} \text{Tr}\left( G J(\eta) G^2 J_{pq} \right)
\end{align}
and 1-chain 
\begin{align}
    ( H^{(0)}\cap \eta)_{pq}
\end{align}
differ by an exact 1-chain if $\delta\eta=0$.

The cap product of 2-chain and 1-cochain is given by
\begin{align}
      ( H^{(0)}\cap \eta)_{pq} = -\frac 1 6 \sum_r  H^{(0)}_{pqr}(\eta(p,r)+\eta(q,r)).
\end{align}
Expanding this expression we find 
\begin{align}
\begin{split}
   ( H^{(0)}\cap \eta)_{pq} = -\frac 1 6 \oint_{z=E_0} \frac{dz}{2\pi i} \text{Tr}\Big( \sum_rGJ_{pq}GQ_r\eta(q,r)+ \sum_rG J_{pq} GQ_r \eta(p,r)\\+2G (J\cap\eta)_q G Q_p+ \sum_rG J_{qr} GQ_p \eta(p,r)-2G (J\cap\eta)_p G Q_q+ \sum_rGJ_{rp}GQ_q \eta(q,r)\Big),
\end{split}
\end{align}
where the cap product of 1-chain and 1-cochain is 
\begin{align}
   ( J\cap\eta)_p = \frac 1 2 \sum_r J_{pr}\eta(p,r)
\end{align}
Two terms in this sum can also be rewritten in this form
\begin{align}
\begin{split}
\sum_r\oint_{z=E_0} \frac{dz}{2\pi i} \text{Tr}\Big( GJ_{rp}GQ_q \eta(q,r) \Big) = \oint_{z=E_0} \frac{dz}{2\pi i} \text{Tr}\Big( - 2G (J\cap\eta)_p G Q_q \Big)
\end{split}
\end{align}
where we have used cocycle condition $\delta \eta = \eta(p,q)+\eta(q,r)+\eta(r,p)=0$ and ultralocallity of the charge $[Q_p,Q_q]=0$. Similarly
\begin{align}
\begin{split}
\sum_r\oint_{z=E_0} \frac{dz}{2\pi i} \text{Tr}\Big( GJ_{qr}GQ_p \eta(p,r) \Big)= \sum_r\oint_{z=E_0} \frac{dz}{2\pi i} \text{Tr}\Big( GJ_{qr}GQ_p  (-\eta(r,q)-\eta(q,p)) \Big) \\=\sum_r\oint_{z=E_0} \frac{dz}{2\pi i} \text{Tr}\Big(GJ_{qr}GQ_p (-\eta(r,q)) \Big) = \oint_{z=E_0} \frac{dz}{2\pi i} \text{Tr}\Big(  2G (J\cap\eta)_q G Q_p \Big)    
\end{split}
\end{align}
The remaining two terms 
\begin{align}
\begin{split}
     \sum_r \oint_{z=E_0} \frac{dz}{2\pi i} \text{Tr}\Big(GJ_{pq}GQ_r\eta(q,r)+G J_{pq} GQ_r \eta(p,r)\Big)
\end{split}
\end{align}
can be rewritten in the same form after the addition of an exact term 
\begin{align*}
 \partial A&=   \sum_r \oint_{z=E_0} \frac{dz}{2\pi i} \text{Tr}\Big( -GJ_{pq}GQ_r (\eta(q,r)+\eta(p,r)) -GJ_{qr}GQ_p (\eta(r,p)+\eta(q,p))\\ &-GJ_{rp}GQ_q (\eta(p,q)+\eta(r,q))\Big) =  \sum_r \oint_{z=E_0} \frac{dz}{2\pi i} \text{Tr}\Big( -GJ_{pq}GQ_r \eta(q,r)-GJ_{pq}GQ_r\eta(p,r)\\& -GJ_{qr}GQ_p \eta(r,p) -GJ_{rp}GQ_q \eta(r,q)\Big)= \oint_{z=E_0} \frac{dz}{2\pi i} \text{Tr}\Big( -\sum_rGJ_{pq}GQ_r \eta(q,r)\\&-\sum_rGJ_{pq}GQ_r\eta(p,r) +2G (J\cap\eta)_q G Q_p- 2G (J\cap\eta)_p G Q_q   \Big)
\end{align*}
where 2-chain A is defined in the first equality. Summing all this together we find that up to an exact term
\begin{align}
    ( H^{(0)}\cap \eta)_{pq} =   \oint_{z=E_0} \frac{dz}{2\pi i} \text{Tr}\Big(G (J\cap\eta)_p G Q_q-G (J\cap\eta)_q G Q_p\Big)+{\rm exact}.
\end{align}
We see that the difference of the Hall conductance and $ H^{(0)}$ is
\begin{align*}
    &\sigma^H(\eta)_{pq}-( H^{(0)}\cap \eta)_{pq} \\&= -i\oint_{z=E_0} \frac{dz}{2\pi i} \text{Tr}\left( G J(\eta) G^2 J_{pq} \right)- \oint_{z=E_0} \frac{dz}{2\pi i} \text{Tr}\Big(G (J\cap\eta)_p G Q_q-G (J\cap\eta)_q G Q_p\Big)\\&=-i\sum_r \oint_{z=E_0} \frac{dz}{2\pi i} \text{Tr}\Big(G (J\cap\eta)_r G^2 J_{pq}+G (J\cap\eta)_p G^2 J_{qr}+G (J\cap\eta)_q G^2 J_{rp}\Big)
\end{align*}
is an exact 1-chain.
\bibliographystyle{apsrev4-1}
\bibliography{bib}
\end{document}